\documentclass[%
 aip,
 jcp,%
 amsmath,amssymb,
reprint,%
]{revtex4-1}

\usepackage{hyperref}
\usepackage{graphicx}
\usepackage{bm}
\usepackage{amssymb,amsmath}
\usepackage{booktabs,array,dcolumn}
\usepackage{siunitx}

\begin{document}

\title{Perspective: Inverse methods for material design}

\author{Avni Jain}
\affiliation{McKetta Department of Chemical Engineering, The University of Texas at Austin, Austin, TX 78712}

\author{Jonathan A. Bollinger}
\affiliation{McKetta Department of Chemical Engineering, The University of Texas at Austin, Austin, TX 78712}

\author{Thomas M. Truskett}
\email{truskett@che.utexas.edu}
\affiliation{McKetta Department of Chemical Engineering, The University of Texas at Austin, Austin, TX 78712}

\date{\today}

\begin{abstract}
 A revised version of this manuscript has been accepted in the AIChE Journal \href{http://onlinelibrary.wiley.com/doi/10.1002/aic.14491/abstract}{[\smash{\underline{website link}}]} and can be cited as doi: 10.1002/aic.14491.
\end{abstract}

\maketitle



\noindent{\textbf{Introduction}}
\vspace{5pt}

\noindent Discovery and design of new materials can be conceptualized via the hierarchy shown in Fig.~1, in which engineering performance derives from dominant structural characteristics exhibited at various length scales.
Structural features in a material can emerge spontaneously via self- or directed-assembly of primary building blocks (molecules, nanoparticles, colloids, etc.), and they can also be imposed using top-down fabrication techniques. 
This hierarchical perspective has been enriched by the widespread use of powerful experimental characterization techniques, which provide micro- to mesoscopic information about the morphologies of
materials with known macroscopic behaviors.
Meanwhile, advances in both simulation methods and computing resources enable the modeling of materials from quantum to continuum scales, offering new opportunities to understand not only how specific structures can form in materials but also how those structures relate to other properties of interest.   
%

Both new materials and new material design rules have been discovered by traversing the hierarchy of Fig.~1 via `forward' strategies, in which (i) samples with diverse structures are examined to understand how morphology impacts a macroscopic property, or (ii) precursors and synthesis parameters are explored combinatorially to explore the structures that are accessible to a system.
That materials with certain periodic structures, such as diamond, can exhibit desirable photonic properties \cite{DIAphoton,ELThomasdiamondNatMAt} or that nanoscale lamellar motifs on the feet of geckos might be mimicked to create dry adhesive materials \cite{geckoNatMat} represent types of findings that have resulted from the first approach.
An example of the second forward strategy is the structural characterization of materials with self-assembling units, such as colloids\cite{Vanmaekelbergh2011419}, quantum dots\cite{ANGEChemTuan2006,QDreview2012,YixuanYu2013CPCSiNC} or metallic nanoparticles \cite{JCPB2003KennethKlabunde,korgelNanoLett2010}, in which the particle composition and other system parameters are varied to tune the kinetics of assembly and the symmetry of the resulting superlattices \cite{MPPileni2011,bian2011shape,royshenhar2012}.

\begin{figure}
  \includegraphics{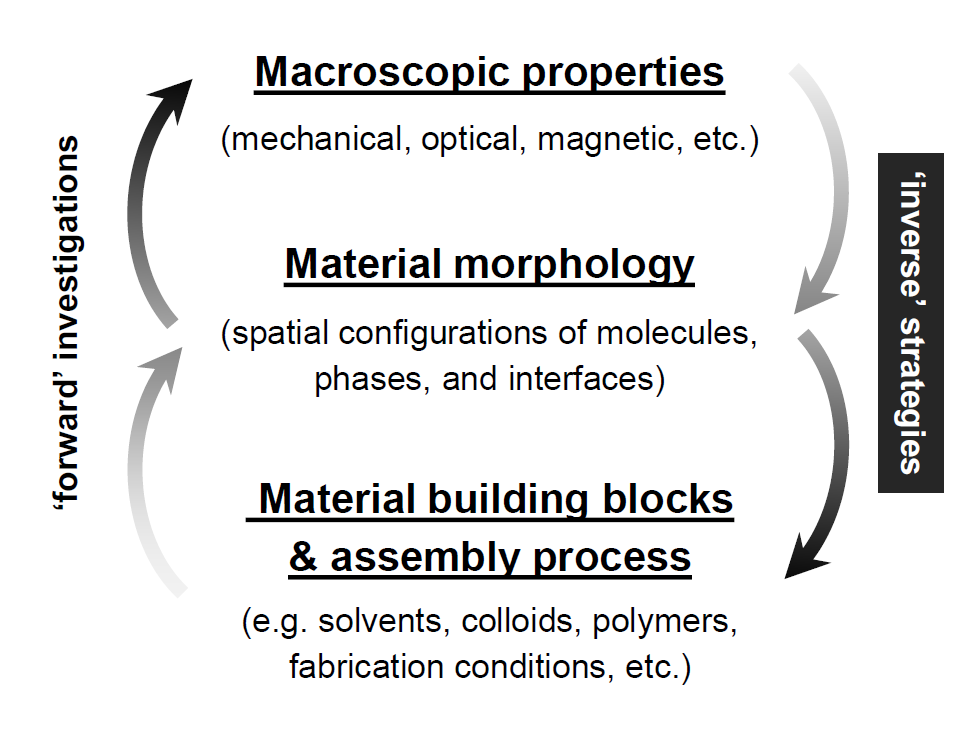}
  \caption{A hierarchical view of material discovery and design.} 
  \label{sch:Figure1}
\end{figure}

From a design perspective, however, it is perhaps most natural to begin with a set of macroscopic properties specified by an application and then proceed down the hierarchy of Fig.~1 using `inverse' strategies to discover which structures, material precursors, and fabrication methods can produce materials consistent with those specifications.
At the coarsest level, such inverse approaches simply ask whether there is a class of morphologies that would be optimal (within known material property constraints) for realizing the specified macroscopic properties. 
Once optimal morphologies are identified, the focus redirects toward how best to design fabrication or assembly routes with appropriate material precursors for realizing the required structures.
%

In this Perspective, we highlight several recent studies that illustrate how inverse strategies using appropriate physical models and computational methods can address the following complex materials design questions within the hierarchical framework of Fig.~1.
\begin{itemize}
\item Which microstructures reflect--or, alternatively, suppress transmission of--target wavelengths of light for design of structural color or photonic band gap materials, respectively?
\item Which `sparse' template structures direct the assembly of block copolymers into target morphologies for graphoepitaxial nanopatterning applications? 
\item Which isotropic interactions between colloidal particles promote their self assembly into targeted, open superlattice structures?
\end{itemize}
We also briefly consider future applications where inverse design methods might return rapid dividends, highlight current limitations of inverse strategies, and speculate on how some of the challenges of this field might be addressed.

\vspace{5pt}

\noindent{\textbf{Designing structures for target properties}}

\vspace{3pt}

\noindent{\emph{Dielectric thin-film morphologies for structural color applications.}} Numerical methods, led by finite-element based topology optimization (TO), have been used since the late 1980s to determine material microstructures that meet design specifications. \cite{TOReviewSigmund2013}
Initially, these methods were used to design heterogeneous or multiphase materials with targeted thermal expansivity\cite{TOcompositethermexp1996,TOthermalexpa1997} as well as elastic \cite{Sigmund1995351} and piezoelectric \cite{TOpiezoelec1998} properties.
During the past decade, however, the TO algorithm has been extended to discover new microstructures consistent with desired acoustic\cite{TOacoustic2008}, nanophotonic\cite{TOnanocolor2014}, and photovoltaic \cite{TOabsorption2013} behaviors as well. 
Non-linear constrained optimizations have also been employed to design surface textures of thin film solar cells for enhanced photon absorption\cite{optimizethinfilmSicells2011} and to discover novel amorphous photonic structures\cite{Torquato2DrandomPNAS2013}.

\begin{figure}
  \includegraphics{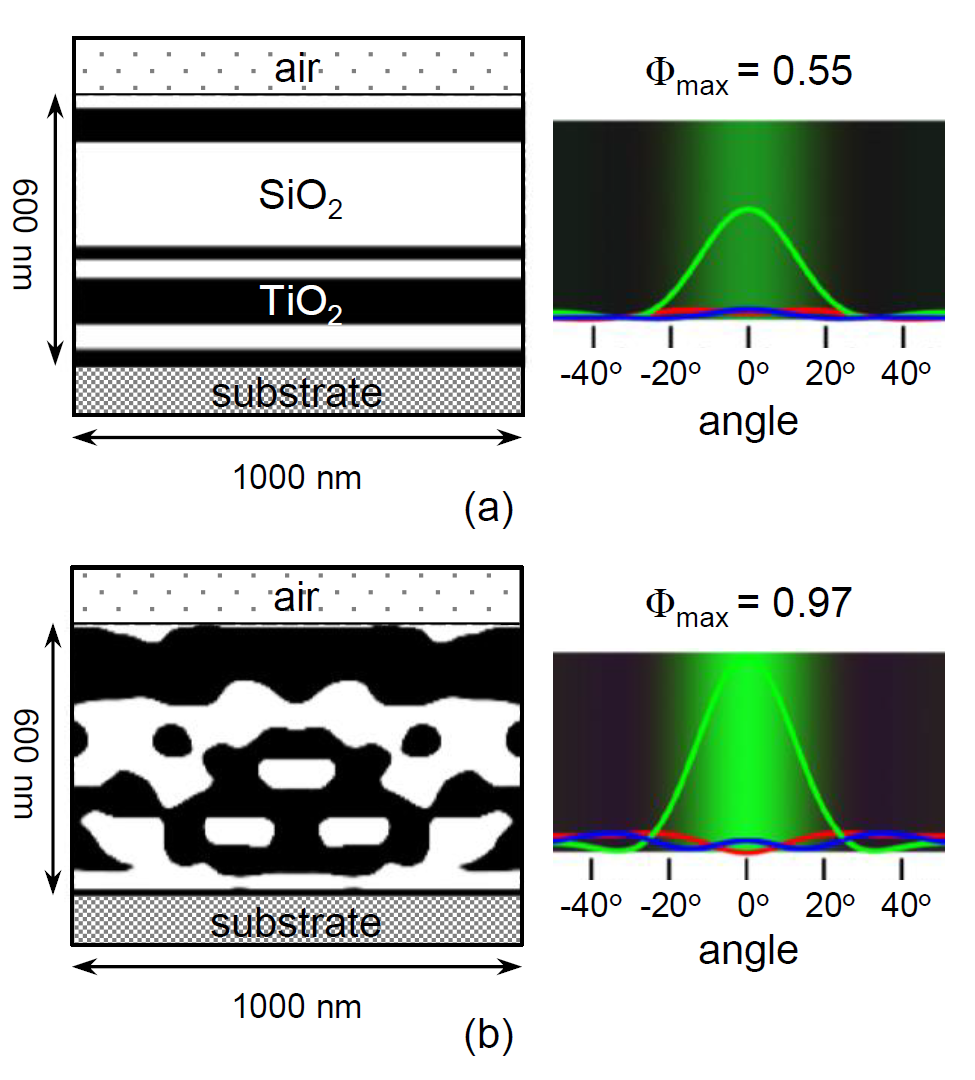}
  \caption{Left panels illustrate silica (\(\text{SiO}_{\text{2}}\)) and titania (\(\text{TiO}_{\text{2}}\)) thin-film morphologies optimized to maximize intensity of a prescribed color (green) in a prescribed direction (0\si{\degree}). Right panels show simulated red-green-blue (RGB) response curves for the designs under exposure to full-spectrum daylight at various incident angles, where \(\Phi_{\text{max}}\) is the maximum intensity. In (a), the optimization was constrained to test only alternately layered phases; in (b), this constraint was relaxed. Adapted with permission from \cite{TOnanocolor2014}.}
  \label{sch:Figure2}
\end{figure}

Inspired by the vibrantly hued yet non-pigmented scales of \emph{Morpho} butterflies\cite{SaitoAbio-photonic,Photonicbiology} and other insects\cite{Whitebeatles}, in which the apparent colors arise from the interaction of light with the wings' nanoscopic cuticle-air features, researchers recently used TO to control perceived color solely by manipulating the material architecture.\cite{TOnanocolor2014}
Specifically, the authors optimized thin-film dielectric morphologies to produce specific and high-intensity structural color responses; i.e., without the use of individual pigments. 
In Fig.~2, the left-hand panels illustrate silica (\(\text{SiO}_{\text{2}}\)) and titania (\(\text{TiO}_{\text{2}}\)) dielectric films with nanostructural features optimized for a green response on exposure to daylight at a specular angle of 0\si{\degree}. 
Two design optimizations were explored. Fig.~2(a) shows the result when the dielectric material was required to be alternately layered (i.e., one-dimensional gratings), and Fig.~2(b) shows the more complex structure of the unconstrained optimum. The right-hand panels provide the red-green-blue (RGB) response curves over a wider angular spectrum with the background displaying the perceived color. 
While both the layered and disordered morphologies meet the set objective, the intensity is clearly optimized by the disorganized microstructure.

More broadly, it was demonstrated that the qualitative structural features of the optimal films depend sensitively on the desired color response in a way that could not be guessed a priori without theoretical guidance. Designs for a wide range of colors were presented, and for some wavelengths even the unconstrained optimizations surprisingly found layered morphologies that were optimal \cite{TOnanocolor2014}. For angle-dependent (iridescent) color responses, entirely different microstructures are required with the details depending on the manufacturing constraints. 
In short, this theoretical strategy, rooted in deeply-established physics principles \cite{Bao:95,bao1995finite,Dossou2006120}, provides an example of how inverse methods can make very specific and non-trivial theoretical predictions about the classes of microstructures that should be tested and possibly pursued in the next stage of experimental design.      
\vspace{5pt}


\begin{figure}
  \includegraphics{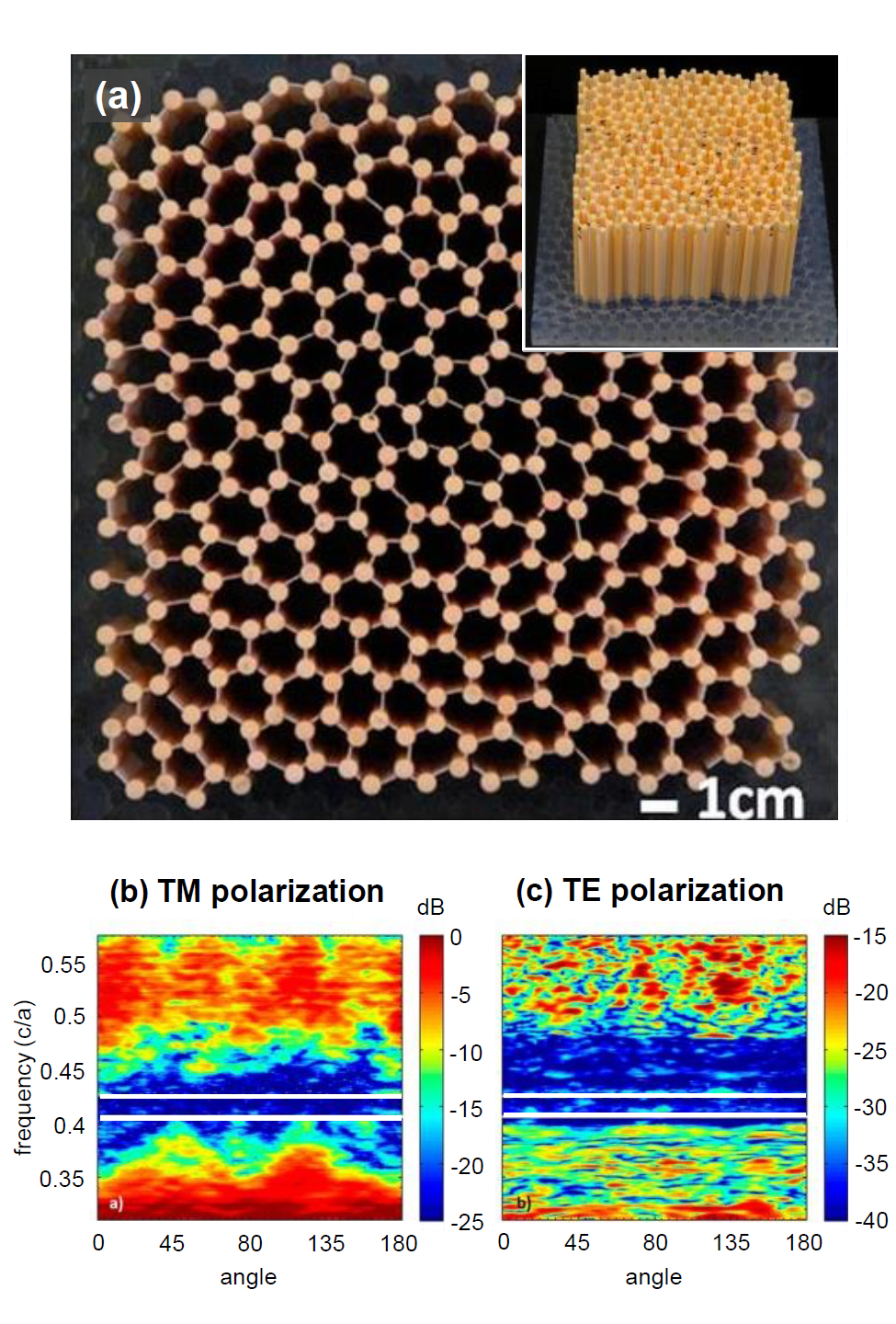}
  \caption{(a) A fabricated amorphous 2D pattern comprising alumina cylinders and connective sheets, designed to display a complete photonic band gap (PBG) (inset shows a side view of the structure). The lower panels show measured transmission strength (in dB) for various light frequencies and incident angles. The PBGs in response to (b) transverse magnetic (TM) and (c) transverse electric (TE) polarizations are bounded by white lines. Adapted with permission from \cite{Torquato2DrandomPNAS2013}.
  }
  \label{sch:Figure3}
\end{figure}

\noindent{\emph{Disordered materials with complete photonic band gaps.}} 
A photonic band gap (PBG) is a frequency range over which transmission of incident electromagnetic waves is blocked.
By combining at least two material components with disparate dielectric constants in specific ratios and configurations, the transmission of light frequencies can be blocked for some or all incident directions and polarizations, the latter corresponding to a `complete' PBG.

Both forward and inverse studies of photonic materials have discovered periodic microstructural configurations that exhibit PBGs\cite{soukoulis2001photonic}, including diamond\cite{DIAphoton} and woodpile structures\cite{ELThomasdiamondNatMAt}, and cylindrical packings in hexagonal\cite{2DPBGmaterialPRB1996} and honeycomb\cite{OptimalPhXtalPRL} patterns. 
Beyond periodic structures, a developing research area considers how light interacts with disordered and nonperiodic structures \cite{Wiersma2013Nat}.
Examples include photonic glasses\cite{Photonicglass}, display of structural colors in animals\cite{Photonicbiology}, and packing of avian photoreceptors \cite{NewPRE2014Aviancells}.
Despite the interest in disordered photonic materials, broad principles for designing disordered PGB microstructures have been slow to emerge.

A recent inverse design study\cite{Torquato2DrandomPNAS2013} focused on discovering novel microstructures consisting of irregular 2D point patterns that exhibit complete PBGs. 
To make the optimization tractable, the authors constrained their design space to `hyperuniform' structures only \cite{PRE2003hyperuniform,PRB20092DquasiPBG}, which exhibit suppressed density fluctuations
(all crystals, quasicrystals and a subset of disordered patterns--e.g., those corresponding to maximally random jammed particle packings \cite{PhysRevLett.84.2064}--are hyperuniform). 
Using pre-generated hyperuniform point patterns to position alumina cylinders, the authors numerically optimized the common cylinder radius to maximize the PBG frequency range.
Fig.~3 illustrates a fabricated version of the optimal 2D pattern and the measured PBGs, which matched the theoretical predictions.
Most importantly, the authors were able to quantify common structural characteristics of optimal hyperuniform 2D structures for PBG materials, including their enhanced short-ranged geometric order (as characterized by the strength of cylinder-cylinder spatial correlations) and the uniformity of their local topology (as measured by the cylinder coordination number). 
The combination of these characteristics--suppressed density fluctuations, short-range correlations, and uniform topology--provides a new structural `design rule' discovered by inverse methods that has more recently been used to generate, using Direct Laser Writing techniques ~\cite{deubel2004direct,2013-DLW,2014-DLW-SiHyperuniform}, synthetic disordered 3D photonic structures yielding complete PBGs \cite{PRA3Dphot}.

\vspace{5pt}


\noindent{\textbf{Designing directed- and self-assembly processes for target structures}} 

\vspace{3pt}
\noindent Once desirable target structures are identified, the challenge of how to synthesize materials with the required morphologies remains. 
Broadly speaking, there are two types of approaches: (1) top-down fabrication and (2) bottom-up assembly.  
Top-down approaches typically use pattern transfer, etching, or deposition technologies to impose the desired structural features on the materials of interest \cite{Patternbook,Biswas2012Rev,Ulissi2013149,topdownfab,ImprintLith,3DPrintRev2014}.
Assembly, on the other hand, relies upon spontaneous formation of target morphologies, which can be driven by `self' interactions between the primary units or building blocks of the material system or additionally `directed' by carefully chosen external fields or boundaries in the system \cite{XiaYPersp2000,EMFurstACSNano2010,RevExtAssemblyNov2013}.
In many applications, both top-down and bottom-up strategies play a role. For example, directed assembly often relies upon top-down fabrication methods like lithography to create an initial template (e.g., a chemically or topographically pre-patterned substrate) that helps to steer the assembly of smaller primary units (e.g., block copolymers) into a desired structure \cite{NP2004APAli,EdThomas2006,NanofabWilsonRev2005,NanosphereLith2013}.

From a fundamental perspective, directed- and self-assembly methods pose challenging questions about how best to choose system parameters and material components in order to promote organization of a condensed matter system into a target structure instead of a competing morphology. 
As we describe below, inverse methods for assembly can take advantage of multidimensional optimization methods and statistical mechanical theories of complex fluids, and thus they are natural tools to help address these questions.
\vspace{5pt}


\noindent{\emph{Templates for directed assembly of block-copolymer (BCP) morphologies.}} 
As a first step, templates for directed assembly are typically designed by intuition: imprinting physical or chemical patches along regions where complementary phases are to be located. 
A key challenge is to use inverse design theory to develop templates with minimum feature density and complexity while still inducing assembly of structures consistent with device-pattern specifications \cite{ForwardCalBCPsparsePattern2008,forwardstudyBCPsparsetemp2010}.
A recent series of studies\cite{KatzInverseAssembly2013,KatzNL2014,chang2014design} used Monte-Carlo (MC) based optimizations to design substrates with maximally sparse patterns that promote target BCP phase-separated patterns, including technologically-relevant features like bends, junctions, and terminations.
The templates consisted of posts selectively attracted to one of the BCP phases situated on a phase-neutral substrate. 
Fig.~4 presents the authors' inverse workflow, in which they (a) specified the target BCP morphology and (b) optimized the post configuration (shown as dots) based on predictions from self-consistent field theory (SCFT). 
They then (c) fabricated the template using electron beam lithography and (d) demonstrated that the target polystyrene-b-polydimethylsiloxane (PS-b-PDMS) BCP morphology was indeed assembled in the optimized post configuration. 
As an alternative to the MC-based inverse strategy, other researchers\cite{JianQinSM2013} designed similar BCP templates using an evolutionary algorithm based on Cahn-Hillard equations for non-equilibrium phase separation. 
While their optimal templates have yet to be validated experimentally, the authors emphasized that using analytical theories should make increasingly complex design problems computationally tractable.
\vspace{5pt}

\begin{figure}
  \includegraphics{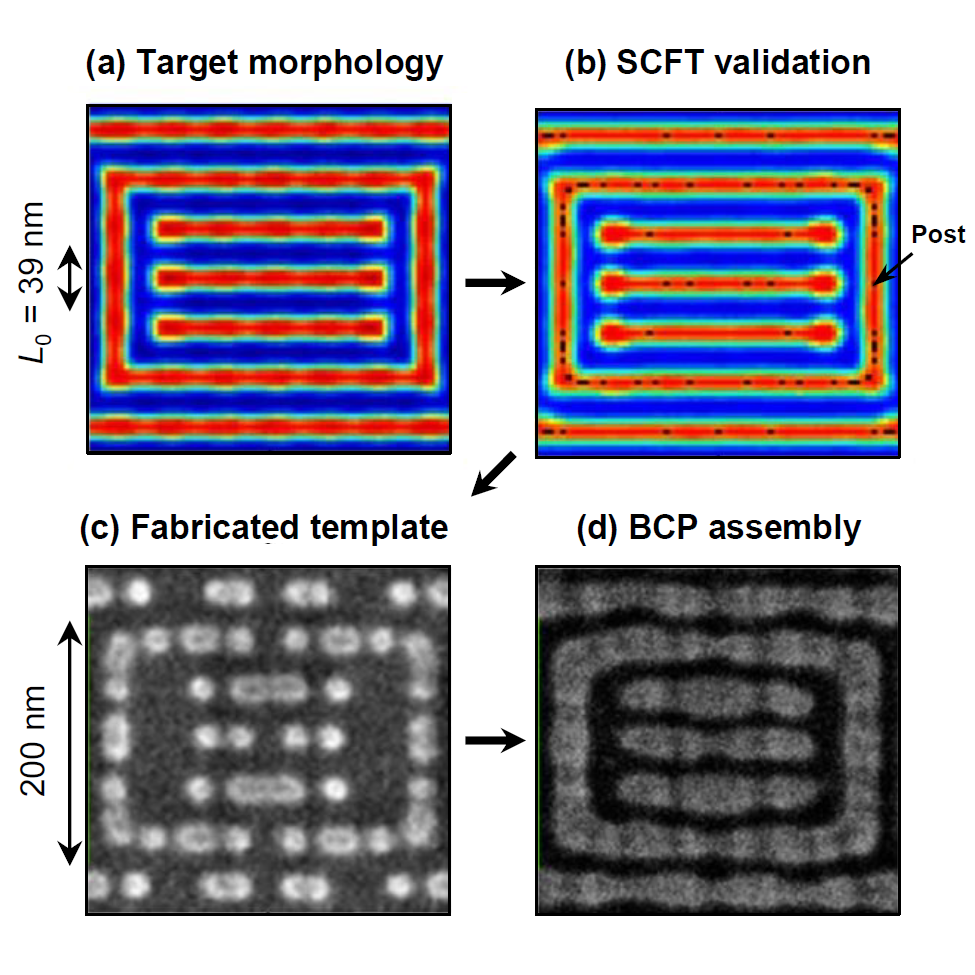}
  \caption{(a) Target polystyrene-b-polydimethylsiloxane (PS-b-PDMS) block-copolymer (BCP) morphology 
  (minority PDMS regions are red, majority PS regions are blue). (b) Simulation results confirming the target structure is displayed using the inversely optimized template of posts (shown as black dots). (c) The optimal template, fabricated via electron beam lithography. (d) SEM scan of the resulting BCP morphology. Adapted with permission from \cite{KatzNL2014}.
  }
  \label{sch:Figure4}
\end{figure}


\noindent{\emph{Designing interactions for targeted superlattice self-assembly.}} 
An important focus of the modern self-assembly literature has been on understanding which equilibrium superlattice phases are formed by various nano- or microscale colloids\cite{glotzer2004self} based on their shape\cite{glotzer2007anisotropy,ExpShapeNP2009,Glotzerscience,EscobedoNatMat2011,PtNanocubes,RevShape2013}, size\cite{murray2000synthesis,shevchenko2006structural,royshenhar2012}, surface texture\cite{SurfRoughPNAS2012}, surface coating\cite{akcora2009anisotropic,ReviewpatchyMay2013,AGP-SK-SM-2014}, particle-solvent interactions\cite{MPPileni2011}, etc.
A particularly fruitful topic has been the examination of assembly driven by particle-surface functionalizations which provide orientation-specific interparticle interactions\cite{patchyGKahl,designrulepatchy_glotzer,sciortinonatcom} that mimic atomic covalent bonding via, e.g., DNA sticky ends \cite{TkachenkoPRE2013,latestOG2013,DNArev2013,MOdC-Nature-DNA-2014} or similar interactions promoted by complementary inorganic ligands \cite{akcora2009anisotropic,WangsciencepineNov2012,Moffitt2013Review}. 

However, outstanding challenges remain in characterizing the effective interactions between suspended nano-(or colloidal) particles as well as in elucidating their stable equilibrium phases. 
A key challenge is to identify the basic limitations that interaction isotropy between particles places on the types of achievable periodic structures. 
This is important because there are practical advantages in terms of the synthesis and assembly kinetics of nano (or colloidal) particles with approximately isotropic (versus orientation-dependent or ``patchy'') interactions\cite{Energylanscapechargedjanus}.

Statistical-mechanics based computational studies \cite{rev2009STInverse,SMJainTMT2013,marcottediamondpaper,edlundjcp2013,ZhangTorquato2013} have demonstrated that isotropic interactions can drive assembly of exotic, low-coordinated 2D (e.g. Kagome, honeycomb, snub-square) and 3D (e.g. diamond, simple cubic) lattices, which represents a break from the conventional understanding that directional interactions are necessary to self-assemble such open lattice structures.
Recently\cite{SMJainTMT2013}, authors have used inverse statistical-mechanical optimization methods to design
pairwise interaction potentials that maximized the phase diagram footprints (i.e. density ranges) over which specific 3D lattice, e.g. diamond, ground states were favored. 
In this study, the potentials were constrained to be short-range and convex-repulsive, features qualitatively similar to measured interactions between particles uniformly grafted with ligands interacting in a solvent medium \cite{VlugtTJH,DVTalapinJACS2010}. 
Using free-energy based MC methods\cite{jainJCP}, the associated superlattice phase diagrams were calculated (see Fig.~5) and target phases based on the designed potentials were found to be robust to variations in the osmotic second-virial coefficient of the interparticle interaction (i.e, solvent quality, analogous to temperature in these systems).

\begin{figure}
  \includegraphics{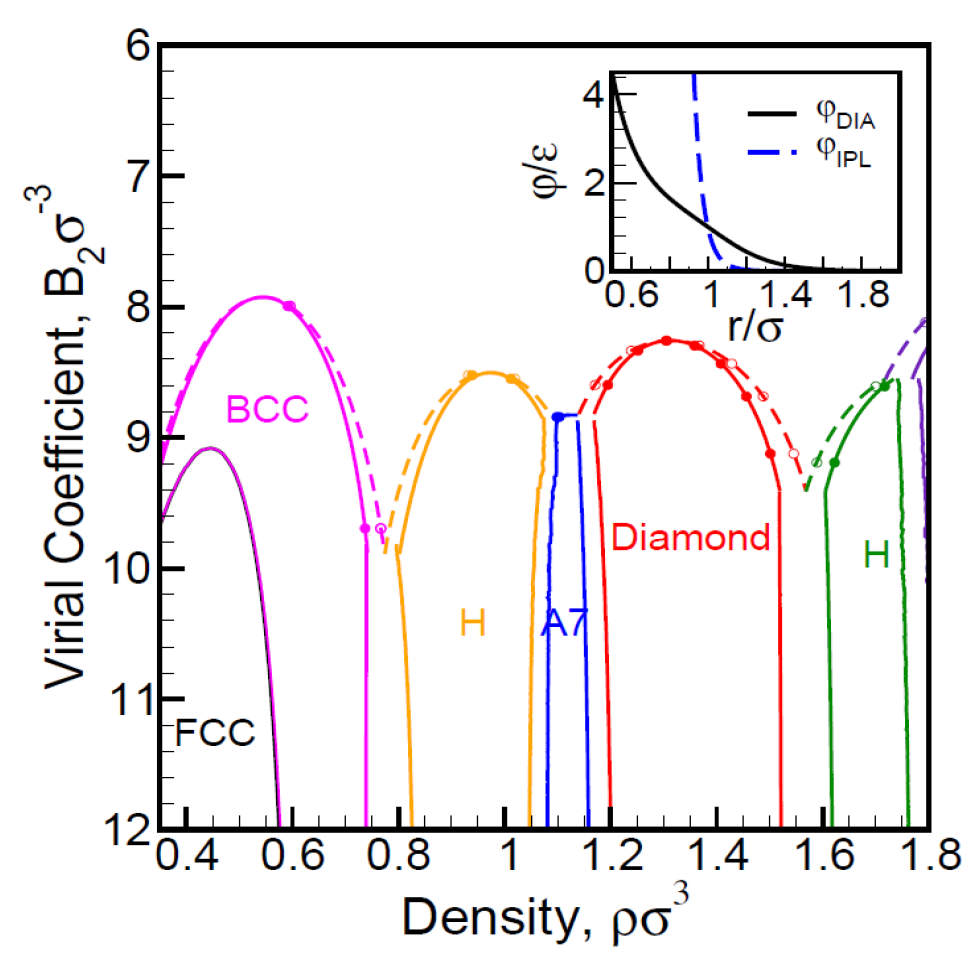}
  \caption{Superlattice-fluid phase diagram of particles with an isotropic interparticle potential optimized to exhibit a diamond crystal ground-state over the widest possible density range. The inset shows the diamond-forming potential (\(\varphi_{\text{DIA}}\)) compared against an inverse power law (n=12) potential (\(\varphi_{\text{IPL}}\)), which highlights the main feature of the optimized potential: its softer repulsion. Coexistence lines were obtained via free-energy MC simulation methods from \cite{FLMC,jctcjeffpaper,jainJCP}.}
  \label{sch:Figure5}
\end{figure}

While these inverse strategies can provide precise mathematical interactions that drive assembly to target structures, realizing these effective potentials in real systems remains an open challenge~\cite{Moffitt2013Review,Fernandes2013Review,LangmuirBockstaller2013}. 
Certainly, experts in colloidal particle synthesis can attempt to creatively intuit the precursor materials and fabrication conditions that result in approximations of optimal effective potentials, though at present this represents a rather open-ended enterprise. 
A more insightful line of inquiry might be to optimize interaction models composed explicitly of experimentally-controllable parameters (e.g. colloidal core material, particle size, grafted ligand lengths, graft densities, etc.).
In one such investigation\cite{PNAS2013inverseDNAgraft}, researchers used genetic algorithms to design the density and length of complementary sticky ssDNA strands to be grafted on spherical colloids to form target crystal structures. 
They based their optimization on a previously established numerical model\cite{chadmirkinscience2011} and thus, were able to replicate experimental observations of crystalline assembly for similar DNA-grafted colloidal systems\cite{chadmirkinscience2011}.
Unfortunately, few existing models capture the overarching physics of colloidal interactions over accessible design spaces while remaining tractable for optimization search routines.\cite{Vincent1986,Zhulina1990495,Currie2003205} 
Overcoming this knowledge gap necessitates a return to theoretical work in complex fluids, focused on establishing accurate analytical and numerically tractable models that reflect nano- and microscale colloidal physics.
\vspace{5pt}


\noindent{\textbf{Future directions}}

\vspace{5pt}

\noindent Beyond the highlighted studies above, there are several emerging technological contexts in which computational inverse strategies are already being used, and others where they might return quick dividends.
These endeavors include studies focused on optimizing 
solar cell domain distributions for increased photovoltaic efficiency \cite{optimizethinfilmSicells2011}; 
mesoscale architectures composed of nanocrystals for enhanced energy flow \cite{FuncMesoACR2014DMill}; 
dopant concentrations in polymer and inorganic matrices for target functionalities \cite{Doped2013ChemMatDMill}; multiphase interfaces for thermoelectric transport \cite{thermoelec2014};
external fields\cite{PRLGaurav} and interparticle interactions\cite{JCSM2012} for enhanced single-particle mobility; and 
faceted colloids that form target periodic structures and tesselations \cite{ATselfassembly2014,Glotzerscience}.
Inverse strategies should also prove useful in optimizing new fabrication approaches, e.g., identifying ideal blade micro-features for controlling thin-film organic-semiconductor crystals at industrial manufacturing scales \cite{Persp2013NatMat,YingDiaoFLUENCENatMAt2013}, and
also offer the opportunity to revisit long-standing physical questions, including the evolution of granular shapes that possess target macroscopic stress responses \cite{NatMat2013inverseGranPack}.

It cannot be overemphasized that along with advances in computing capabilities and numerical tools\cite{AIChePerspJJPablo,IEGrossmann2004Persp,Hartke2011,Rev2012Opt,Pasz2013}, the utility of inverse strategies relies on the physical accuracy of the underlying hierarchical connections between properties, structure, and assembly.
As illustrated in this Perspective, many links between macroscopic properties, e.g. mechanical and photonic properties, and underlying microstructures are well understood and have been exploited to design new materials and establish new design rules. 
Also, as device properties are increasingly understood to emerge from material attributes at the nanoscale and smaller, continuum and classical descriptions of materials may reach their fundamental thresholds of applicability. 
Going forward, their validity needs to be established across multiple lengthscales, and, if necessary, new integrated multiscale connections must be developed.
Such challenges are already apparent in terms of holistically understanding the physics of assembly, as exemplified in our discussion of the inverse design of self-assembling colloids.
And while significant knowledge gains are being made through extensive forward-strategy investigations, their successful integration into inverse methods depends on distilling and validating computationally tractable physical theories.
%

We note that the hierarchy of Fig.~1 is a coarse simplification, meant to encompass the broadest possible collection of contemporary material design problems.
It may not always be reasonable or advantageous to pose inverse design problems that separately address either Property \(\rightarrow\) Structure \emph{or} Structure \(\rightarrow\) Assembly connections in isolation.
Especially when macroscopic material properties can be related to the chemical make-up of the material precursors, it may be possible to tune the precursors directly for the desired property, without explicitly specifying structural connections.

The physical models and problem constraints, i.e. `user inputs', that must be chosen while formulating the inverse strategies are ultimately validated when the outcomes of experimental synthesis and characterization match promising solutions.
When outcomes do not match the theoretical predictions, however, there are still gains to be made.
Forward investigations--whether conducted via experiments or simulations--can be used to clarify system responses within regions of the design space where predictions failed, and the underlying theoretical connections can be revised in light of new experimental findings.
This feedback loop between theory and experiments can iteratively provide novel and non-intuitive physical insights that bridge different levels of the material design hierarchy. 
Subsequently, these new connections can streamline the development of an expanding portfolio of technologically-relevant materials, from desired device properties down to the choice of modular and inexpensive material precursors.

\vspace{5pt}

\noindent{\textbf{Acknowledgments}}
\vspace{5pt}

\noindent The authors gratefully acknowledge support from the Robert A. Welch Foundation (F-1696), the National Science Foundation (CBET-1065357), and the Gulf of Mexico Research Initiative.

\vspace{5pt}

\bibliography{PhaseDiagram_bib,nanocrystal,ao_yukawa}

\end{document}